\documentclass[aps,pra,superscriptaddress,twocolumn,10pt]{revtex4-2}
\usepackage[utf8]{inputenc} 
\usepackage{graphicx}	
\usepackage{bm} 	
\usepackage{amsmath}
\usepackage{amsfonts}
\usepackage{bbold}
\def\e{\mathrm{e}}
\def\ii{\mathrm{i}}
\usepackage{braket}
\usepackage[pdftex,dvipsnames]{xcolor}
\usepackage{hyperref}
\hypersetup{
    colorlinks,
    citecolor=black,
    filecolor=black,
    linkcolor=black,
    urlcolor=black
}

\begin{document}

\title{Phase-stable source of high-quality three-photon \\ polarization entanglement by cascaded downconversion}

\author{Zachary M.E. Chaisson}
\affiliation{Département de physique et d’astronomie, Université de Moncton, Moncton, New Brunswick E1A 3E9, Canada}
\author{Patrick F. Poitras}
\affiliation{Département de physique et d’astronomie, Université de Moncton, Moncton, New Brunswick E1A 3E9, Canada}
\author{Micaël Richard}
\affiliation{Département de physique et d’astronomie, Université de Moncton, Moncton, New Brunswick E1A 3E9, Canada}
\author{Yannick Castonguay-Page}
\affiliation{Department of Physics and Astronomy, McMaster University, Hamilton, Ontario L8S 4L8, Canada}
\affiliation{Département de physique et d’astronomie, Université de Moncton, Moncton, New Brunswick E1A 3E9, Canada}
\author{Paul-Henry Glinel}
\affiliation{Département de physique et d’astronomie, Université de Moncton, Moncton, New Brunswick E1A 3E9, Canada}
\author{Véronique Landry}
\affiliation{Département de physique et d’astronomie, Université de Moncton, Moncton, New Brunswick E1A 3E9, Canada}
\author{Deny R. Hamel}
\affiliation{Département de physique et d’astronomie, Université de Moncton, Moncton, New Brunswick E1A 3E9, Canada}

\begin{abstract}
Stable sources of entangled photons are important requirements for quantum communications. 
In recent years, cascaded downconversion has been demonstrated as an effective method of directly producing three-photon entanglement. 
However,  to produce polarization entanglement these sources have until now relied on intricate active phase stabilization schemes, thus limiting their robustness and usability. 
In this work, we present a completely phase-stable source of three-photon entanglement in the polarization degree of freedom.
With this source, which is based on a cascade of two pair sources based on Sagnac configurations, we produce states with over 96\% fidelity with an ideal GHZ state. 
Moreover, we demonstrate the stability of the source over several days without any ongoing optimization. 
We expect this source to be a useful tool for applications requiring multiphoton entanglement, such as quantum secret sharing and producing heralded entangled photon pairs. 
\end{abstract}

\maketitle

\section{Introduction}

Multi-photon entanglement is an important resource for a wide range of quantum information applications~\cite{pan_multiphoton_2012}. 
Photonic Greenberger-Horn-Zeilinger (GHZ) states in particular are known to be useful for tasks such as quantum secret sharing~\cite{hillery_secret_1999}, quantum anonymous transfer~\cite{christandl_anonymous_2005} and optical quantum computing~\cite{browne_computation_2005}. 
Currently, multi-photon entangled states are most often produced by combining two or more entangled photon pairs from spontaneous parametric downconversion (SPDC) and using post-selection to project onto the desired states~\cite{bouwmeester_GHZ_1999,Pan2001,Eibl2003,Eibl2004,Zhao2004,Walther2005,Lu2007,Yao2012,Wang2016,Chen2017,pilnyak_simple_2017,zhong_twelve_2018}. In this approach, post-selection is fundamental to the state creation process, as photons must first be detected in order to produce the desired entangled state.

An alternative to these post-selection based methods is to instead cascade multiple SPDC sources (C-SPDC) to directly create the desired state~\cite{greenberger_bells_1990}, removing the fundamental requirement of post-selection. 
This novel approach has already been successfully employed  to  produce photon triplets using separate sources~\cite{hubel_direct_2010,ding_hybrid-cascaded_2015} as well as with a cascade within a single integrated device~\cite{krapick_-chip_2016}.

C-SDPC has also been used to produce polarization entanglement and to herald Bell states~\cite{hamel_direct_2014}.
However, demonstrations of polarization-entanglement using C-SPDC have, until now, used Mach-Zehnder interferometer configurations. 
This type of configuration has the advantage of only using the crystals in a single direction, which allows for the use of waveguided crystals pig-tailed with single mode fibers optimized for the pump at the entrance and the downconverted signal at the output. 
However, it also has the significant drawback of requiring active stabilization of the phase between the crystals in each arm, which adds significant complexity to the setup and reduces its robustness in real world application. 
An alternative to the Mach-Zehnder configuration is to instead employ a Sagnac interferometer, removing the need for active stabilization but losing the advantage of optimized single mode fibers. 

In this work, we present a phase-stable source of polarization-entangled photon triplets based on C-SPDC. 
By cascading two sources built using a Sagnac interferometer~\cite{kim_phase-stable_2006, fedrizzi_wavelength-tunable_2007}, which are inherently phase stable, we construct a source which can display high state fidelity with GHZ states without active stabilization.  

\section{Experimental setup}
	
The state we aim to produce through C-SPDC is the three-photon Greenberger-Horn-Zeilinger (GHZ) state, given by :

\begin{equation}
\ket{\mathrm{GHZ}} = \frac{1}{\sqrt{2}} (\ket{HHH} + \ket{VVV})
\label{eq:GHZ}
\end{equation}
	
\noindent where $\ket{HHH}$ represents three photons with horizontal polarization, while $\ket{VVV}$ represents vertical polarization. 
	
\begin{figure}[t!]
\centering
\includegraphics[width=7in]{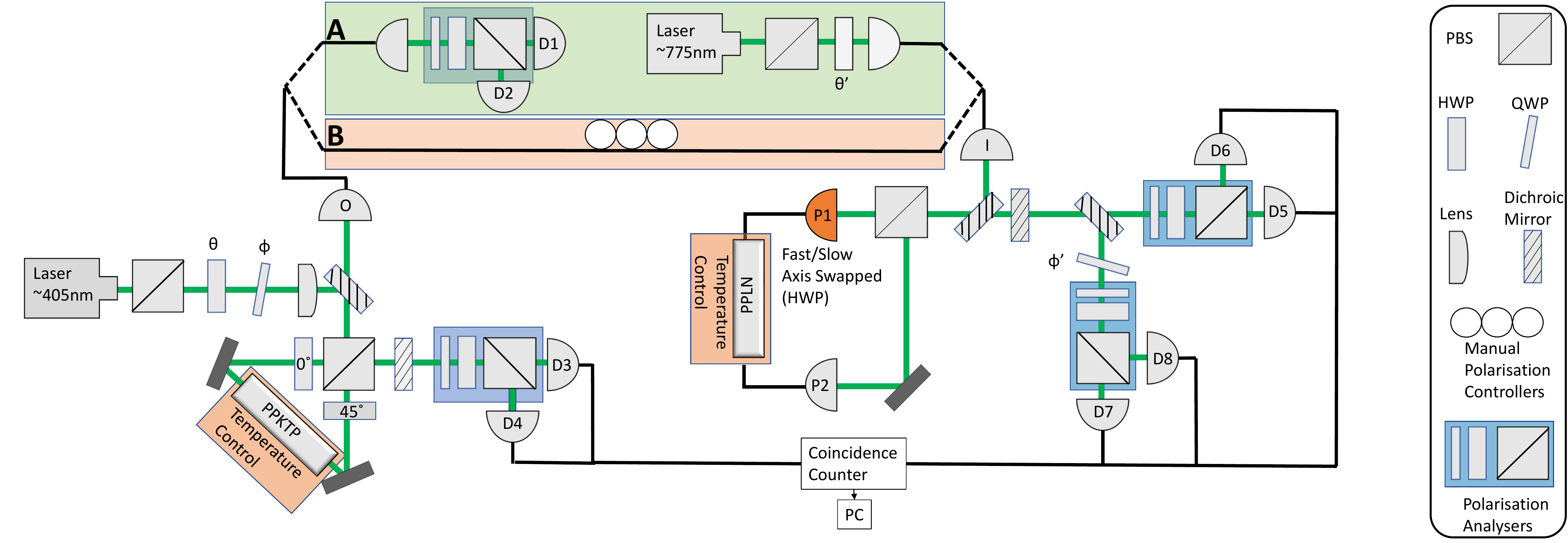}
\caption{Experimental set up used to create and measure entangled photon pairs (\textbf{A}) or triplets  (\textbf{B}). 
In the first source, a 405~nm grating-stabilized laser diode (Toptica Topmode) pumps a PPKTP crystal (Raicol), which is heated at 48.0 $^{\circ}$C to produce photons at 777 nm and 846 nm. The PPKTP crystal is placed inside a Sagnac interferometer, along with superachromatic half-wave plates (Thorlabs SAHWP05M-700). After filtering out the pump with dichroic mirrors, the resulting state is that of Eq.~\ref{eq:Psi1}.
The 846 nm photon is sent to polarization analyzers, whereas the 777 nm photon is coupled into a single mode fiber by collimator O and is either measured directly (\textbf{A}) or sent to the second source after passing through a manual polarization controller (\textbf{B}).
In the second source, a PPLN waveguide is pumped either directly by a wavelength-tuneable grating-stabilized laser diode (Sacher Lynx TEC 150, Littrow Series) to produce the state in Eq.\ref{eq:Phi1} (\textbf{A}), or by 777 nm single photons from the first source to produce Eq.~\ref{eq:GHZexp} (\textbf{B}). 
The PPLN waveguide is heated at 50.0$^{\circ}$C to produce photon pairs at approximately 1530 nm and 1570 nm. 
The fast and slow axis of the collimator P1 (in orange) are swapped to mimic the effects of a half-wave plate at 45 degrees. All photons are detected using superconducting nanowire single-photon detectors (SNSPDs, Photon Spot).}
\label{Montage:CSPDC}
\end{figure}

We start by creating two independent SPDC sources, as seen in Fig. \ref{Montage:CSPDC} (A). 
In the first source, a periodically poled potassium triphosphate (PPKTP) crystal is placed in a Sagnac interferometer. 
This source uses a type-II SPDC process to create states of the form :

\begin{equation}
\ket{\Psi^{\pm}} = \cos \theta \ket{HV} +\e^{\ii \phi} \sin \theta \ket{VH}
\label{eq:Psi1}
\end{equation}

\noindent	where $\theta$ and $\phi$ are determined by the polarization of the pump. These two parameters are controlled by turning a half-wave plate (HWP) and tilting a quarter-wave plate (QWP), respectively.

For the second source, a  type-0 periodically poled lithium niobate (PPLN) crystal waveguide, pig-tailed at each end with polarization maintaining fibers, is placed in a Sagnac interferometer.
Since the crystal is pumped from both directions, it is not possible to use fibers optimized for the pump wavelength.

\onecolumngrid

\twocolumngrid	

Instead, we employ polarization-maintaining fibers which are single mode for the downconverted photons~\cite{lim_stable_2008}, with FC/APC connectors to avoid back reflections. 
The state we aim to produce has the form :
	
\begin{equation}
\ket{\Phi^{\pm}} = \cos \theta' \ket{HH} + \e^{\ii \phi '}  \sin \theta'  \ket{VV}
\label{eq:Phi1}
\end{equation}

Where again the weighing of the terms $\theta '$ and phase  $\phi '$ can be set respectively by turning a HWP and tilting a QWP, although in this source the phase is set by acting on the downconverted photons rather than on the pump.

The two sources are combined by using one of the photons from the PPKTP source as a pump for the PPLN source, as seen in Fig. \ref{Montage:CSPDC} (\textbf{B}). 
A photon in the state $\ket{H}$ is downconverted into the state $\ket{VV}$ while the mode $\ket{V}$ is downconverted to $\ket{HH}$. The resulting state that we obtain is of the form : 

\begin{equation} \begin{split}
\ket{\mathrm{GHZ}}_{\mathrm{Exp}} =  \cos \theta \ket{HHH} +  \e^{\ii \Phi (\phi , \phi ')} \sin \theta  \ket{VVV}
\label{eq:GHZexp}
\end{split}	\end{equation} 

\section{State preparation}

To prepare the desired state, we start with separated sources as shown in Fig. \ref{Montage:CSPDC} (\textbf{A}). This allows us to first optimize  each pair source separately.

The two sources are then connected, as shown in Fig. \ref{Montage:CSPDC} (\textbf{B}), to form the cascaded source. The manual polarization controller is adjusted so that the horizontal and vertical polarizations are conserved between collimators O and I.

The angle of the HWP $\theta$ is set to produce $\ket{HHH}$ and $\ket{VVV}$ with equal probability. This angle is calculated to compensate for any imbalance in coupling efficiency between photons traveling clockwise and counter-clockwise in the Sagnac loops, as measured from photon pairs.

\begin{figure}[h!]
\centering
\includegraphics[width=8.6cm]{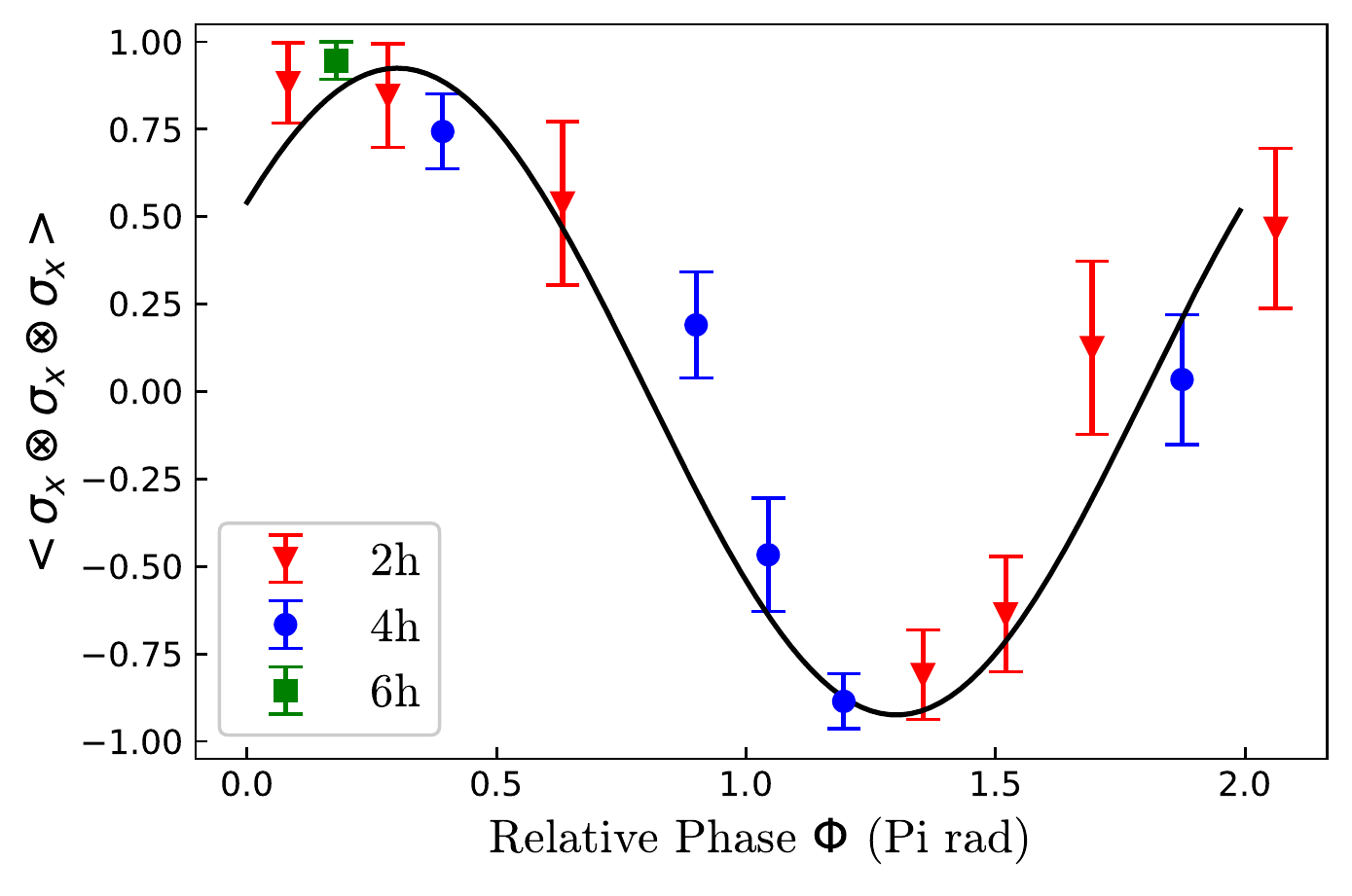}
\caption{Determination of the phase necessary to create the desired GHZ state. By tilting the QWP we can change the relative phase of our state. This is reflected in the measured value (shown here) of the expectation value of the Pauli X triplet measurement. The total length of each measurement is displayed in the legend. Triplet coincidences are detected at a rate of approximately 10 per hour. The black line is a sinusoidal fit, where the phase and amplitude are left as fitting parameters. The fit has an amplitude of $0.92 \pm 0.06$. }
\label{Res:PhasevsAngle}
\end{figure}

With the balance set, we can focus on controlling the phase of the state.
We start by performing a $\sigma_{x} \otimes \sigma_{x}  \otimes \sigma_{x}$ measurement on the photon triplets, where $\sigma_{x}$ is the Pauli X matrix, while varying the phase $\phi$.
These results are given in Fig. \ref{Res:PhasevsAngle}.
As expected, we find a sinusoidal dependence for  $\langle\sigma_{x} \otimes \sigma_{x}  \otimes \sigma_{x} \rangle $, with the fit having a visibility of $0.92 \pm 0.06$.

\section{Results}

\begin{figure}[h!]
\centering
\includegraphics[width=8.6cm]{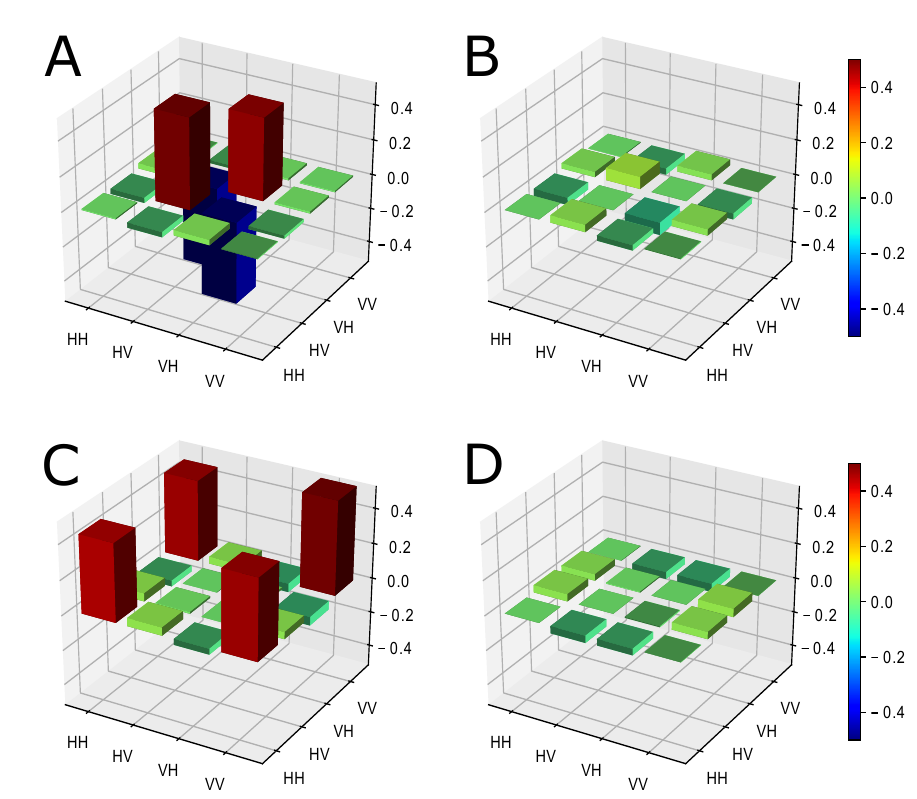}
\caption{Density matrix describing the state of polarization of both sources of entangled photon pairs. For the PPKTP source, the (\textbf{A}) real part and the (\textbf{B}) imaginary part of the density matrix. With a coincidence window of 0.5 ns, the resulting state has a $\ket{\Psi^{-}}$ fidelity of $96.45 \pm 0.01\%$, a purity of $95.61 \pm 0.03\%$ and a tangle~\cite{coffman_distributed_2000} of $91.47 \pm 0.05\%$. For the PPLN source, we have the (\textbf{C}) real and (\textbf{D}) imaginary part of the density matrix, which was measured with a coincidence window of 0.3 ns. The resulting state has a $\ket{\Phi^{+}}$ fidelity of $95.06 \pm 0.05 \%$, a purity of $93.7 \pm 0.1\%$ and a tangle of $86.6 \pm 0.2\%$.}
\label{fig:PairResults}
\end{figure}

We first characterize the pair sources independently.
Using maximum likelihood quantum state tomography~\cite{james_measurement_2001}, we reconstruct the density matrices of the two entangled photon pairs from both sources.
Photons in each output mode are projected onto one of three mutually unbiased polarization bases, (horizontal and vertical, circular right and circular left, diagonal right and diagonal left) for a total of six different polarization measurements per output mode.
For pairs, this leads to a combination of 36 coincidence measurements.
The results of the tomography are shown in Fig. \ref{fig:PairResults}. 

The average dark counts are determined by turning off the pump lasers and is measured at an average of approximately five counts per second on each detector channel.
The twofold coincidences measured at the detectors from the PPKTP source number $3 \cdot 10^{6}$ per second with a pump power of 9.4 mW at the entrance of the interferometer. 
From the tomography, we find the source produces a $\ket{\Psi^-}$ state with a fidelity of $96.45 \pm 0.01$\%.
The PPLN source produces $1.5 \cdot 10^{4}$ twofold coincidences per second with a pump power of approximately 1 uW at the entrance of the crystal.
From the tomography we find that the source produces a $\ket{\Phi^+}$ state with a fidelity of $95.06 \pm 0.05\%$. 
From the ratio of single photon detections to twofold coincidence detections, an average of the combined coupling and detection efficiencies are determined to be 0.30, 0.16 and 0.13 for the 845nm, 1530nm and 1570nm photons, respectively. 
Coupling efficiencies from collimator I to collimators P1 and P2 are measured at 0.30 from single photon detection rates.

For the cascaded photon source, we obtain triplet coincidence rates of approximately 10 per hour.
With a three-qubit tomography requiring measurements from 27 different bases, obtaining the counts for a reconstruction of the density matrix is not feasible in a reasonable amount of time, especially if we want to quantify the stability of the source.
Instead we employ a GHZ witness~\cite{toth_detecting_2005} which requires a measurement in just two basis and gives a lower bound on the fidelity of our state~\cite{guhne_entanglement_2009}. 
The witness is used for the GHZ state given in equation \ref{eq:GHZ}, and is given by :

\begin{equation} \begin{split}
 W_{\mathrm{GHZ}} = &\frac{3}{2} \cdot \mathbb{1}^{3} - 
\sigma_{x} \otimes \sigma_{x}  \otimes \sigma_{x} \\ 
& - \frac{1}{2} \left(  
\mathbb{1} \otimes \sigma_{z} \otimes \sigma_{z} +
\sigma_{z} \otimes \mathbb{1} \otimes \sigma_{z} + 
\sigma_{z} \otimes \sigma_{z} \otimes \mathbb{1}  
\right)
\end{split} \end{equation}
	
\noindent where the $\sigma$ represent their respective Pauli matrices. The lower bound of the fidelity between our experimental state and the GHZ state is given by :
	
\begin{equation}
F_{\mathrm{GHZ}} \geq \frac{1 - W_{\mathrm{GHZ}}}{2}
\end{equation}

\noindent This witness is convenient as it only requires measurements in two measurement bases to obtain a lower bound on the fidelity of our state.

The measurements for the witness were taken over sixteen hours, with eight hours for each basis.
58 three-fold coincidences were measured in the $\sigma_{z}$ basis and 44 in the $\sigma_{x}$ basis for a total of 102 coincidences in 16 hours.
This gives an average of $6.4 \pm 0.6$ triplets per hour.
Results of this measurement are shown in Table \ref{tab:Witness}.
The entanglement witness is violated convincingly with a value of $W_{\mathrm{GHZ}} = -0.92 \pm 0.10$, confirming the entanglement of the state.
This gives a minimum fidelity with the targeted GHZ state of $F_{\mathrm{GHZ}} = 0.96 \pm  0.05$. To the best of our knowledge, this is the highest fidelity reported for a three-photon GHZ state.

\begin{table}[htb]
\caption{Calculated witness results. The calculated errors are one standard deviation, calculated by assuming Poisson noise on the triplet count rate.}
  \begin{center}
    \begin{tabular}{c c c}
    \hline \hline
      \textbf{Measurement} & \textbf{Value} & \hspace*{0.5cm} \textbf{Error}\\
      \hline
      $\sigma_{x} \otimes \sigma_{x}  \otimes \sigma_{x}$ & 0.95 &\hspace*{0.5cm} 0.05\\
      $\mathbb{1} \otimes \sigma_{z} \otimes \sigma_{z}$ & 0.97 &\hspace*{0.5cm} 0.03\\
      $\sigma_{z} \otimes \mathbb{1} \otimes \sigma_{z}$ & 1.00 &\hspace*{0.5cm} 0.04\\
      $\sigma_{z} \otimes \sigma_{z} \otimes \mathbb{1}$ & 0.97 &\hspace*{0.5cm} 0.03\\
      \hline
      $W_{\mathrm{GHZ}}$ & -0.92 &\hspace*{0.5cm} 0.10\\
      Lower bound of $F_{\mathrm{GHZ}}$ & 0.96 &\hspace*{0.5cm} 0.05\\
      \hline \hline
    \end{tabular}
    \label{tab:Witness}
  \end{center}
\end{table}

In order to characterize the stability of the source, the entanglement witness was measured repeatedly over a period of several days.
During this time, no adjustments were made to the source.
As shown in Fig. 4, while the fidelity of the state does display variations, we find that the fidelity stays above $72 \%$ during the entire week-long measurement, with an average fidelity of  $84 \% \pm 8 \%$, indicating that the source has potential for passive long term stability. 

\begin{figure}[h!]
\centering
\includegraphics[width=8.6cm]{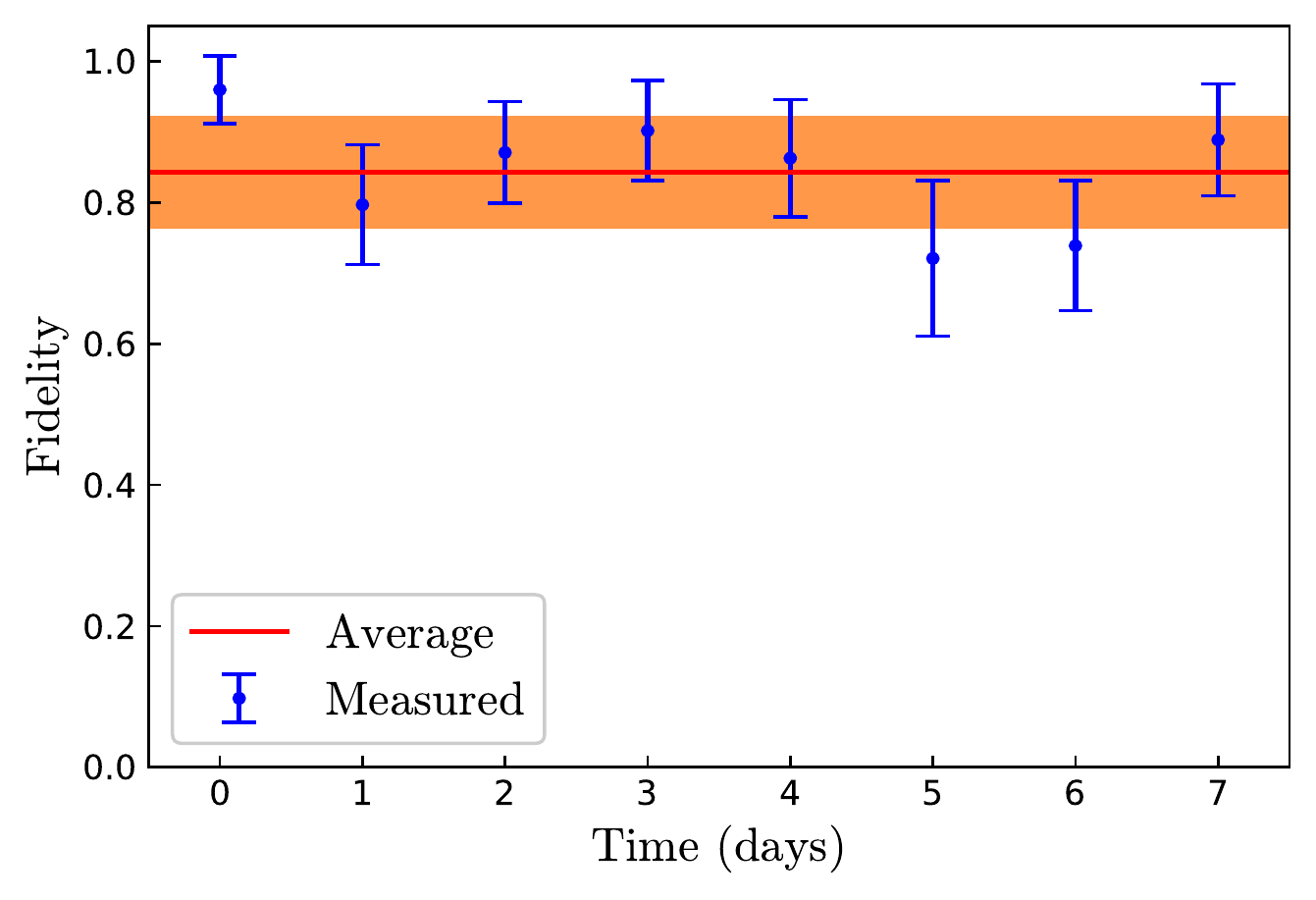}
\caption{Calculated lower bounds of fidelity over time. No adjustments were made to the source once the measurements began. Each of the two Pauli measurements lasted eight hours, followed by an eight hour downtime. The lower bound of the state fidelity begins at a maximum of  $96 \pm 5 \%$. Over the entire seven day period, the average fidelity is  $84 \pm 8 \%$.}
\label{Res:PhasevsTime}
\end{figure}

\section{Discussion}

In terms of source quality, the short term fidelity of  $96~\%$ is excellent. Further improvement efforts should therefore be primarily focused on improving stability and production rates.

While further investigation is required to precisely isolate the source of instability, it is likely that the fibers at each end of the PPLN waveguide are partially responsible, due to the wavelength mismatch with the 775 nm photons. 
It is therefore likely that the wavelength instability could be improved through better thermal stabilisation of this portion of the setup, or perhaps by pig-tailing the PPLN waveguide with endlessly single-mode photonic crystal fibers~\cite{Birks1997}, which would allow for single mode operation at both 775 nm and 1550 nm.

As for the production rates, while the rates for C-SPDC are expected to be low, the experiment had additional limiting factors affecting the rates which are not inherent to the scheme. 
Indeed, previous experiments demonstrated entangled production rates of up to 2 orders of magnitude higher than this work~\cite{hamel_direct_2014}.
The difference is partially due to an additional coupling factor from I to P1 and P2 in Fig. \ref{Montage:CSPDC}, which lowers the expected triplet count rates by a factor of three.
This loss could eventually be avoided by replacing the fiber from O to I with freespace.
A lower pump power in the first source, required to prevent damage to the super-achromatic half-wave plates, accounts for a further factor of two.
A novel approach to the Sagnac interferometer \cite{lee_sagnac-type_2021}, removing the need for achromatic optics, could allow for higher pump intensities. 
The remaining difference in count rates are due to different coupling efficiencies to detectors 5 through 8.
Importantly, none of these additional losses are fundamental to the current scheme, and could therefore be addressed with appropriate improvements to the setup.
Alternatively, count rates could also be improved through the use of non-linear crystals with higher conversion efficiencies~\cite{Wang2018}.
 
With these further improvements to triplet production rates and stability, we expect this source to be of significant usefulness for applications requiring high-fidelity entangled photon triplets. 
In contrast to previous experiments creating three-photon polarization entanglement, our implementation does not rely on post-selection, nor does it require active stabilization, thereby greatly reducing the complexity of the setup and making it attractive for applications requiring robust high-quality entanglement.

\section{Acknowledgments}
We are grateful for the financial support from the Natural Sciences and Engineering Research Council of Canada,  the Canadian Foundation for Innovation, Canada Research Chairs, and the New Brunswick Innovation Foundation. 


%


\begin{thebibliography}{31}%
\makeatletter
\providecommand \@ifxundefined [1]{%
 \@ifx{#1\undefined}
}%
\providecommand \@ifnum [1]{%
 \ifnum #1\expandafter \@firstoftwo
 \else \expandafter \@secondoftwo
 \fi
}%
\providecommand \@ifx [1]{%
 \ifx #1\expandafter \@firstoftwo
 \else \expandafter \@secondoftwo
 \fi
}%
\providecommand \natexlab [1]{#1}%
\providecommand \enquote  [1]{``#1''}%
\providecommand \bibnamefont  [1]{#1}%
\providecommand \bibfnamefont [1]{#1}%
\providecommand \citenamefont [1]{#1}%
\providecommand \href@noop [0]{\@secondoftwo}%
\providecommand \href [0]{\begingroup \@sanitize@url \@href}%
\providecommand \@href[1]{\@@startlink{#1}\@@href}%
\providecommand \@@href[1]{\endgroup#1\@@endlink}%
\providecommand \@sanitize@url [0]{\catcode `\\12\catcode `\$12\catcode
  `\&12\catcode `\#12\catcode `\^12\catcode `\_12\catcode `\%12\relax}%
\providecommand \@@startlink[1]{}%
\providecommand \@@endlink[0]{}%
\providecommand \url  [0]{\begingroup\@sanitize@url \@url }%
\providecommand \@url [1]{\endgroup\@href {#1}{\urlprefix }}%
\providecommand \urlprefix  [0]{URL }%
\providecommand \Eprint [0]{\href }%
\providecommand \doibase [0]{https://doi.org/}%
\providecommand \selectlanguage [0]{\@gobble}%
\providecommand \bibinfo  [0]{\@secondoftwo}%
\providecommand \bibfield  [0]{\@secondoftwo}%
\providecommand \translation [1]{[#1]}%
\providecommand \BibitemOpen [0]{}%
\providecommand \bibitemStop [0]{}%
\providecommand \bibitemNoStop [0]{.\EOS\space}%
\providecommand \EOS [0]{\spacefactor3000\relax}%
\providecommand \BibitemShut  [1]{\csname bibitem#1\endcsname}%
\let\auto@bib@innerbib\@empty
\bibitem [{\citenamefont {Pan}\ \emph {et~al.}(2012)\citenamefont {Pan},
  \citenamefont {Chen}, \citenamefont {Lu}, \citenamefont {Weinfurter},
  \citenamefont {Zeilinger},\ and\ \citenamefont
  {{\.Z}ukowski}}]{pan_multiphoton_2012}%
  \BibitemOpen
  \bibfield  {author} {\bibinfo {author} {\bibfnamefont {J.-W.}\ \bibnamefont
  {Pan}}, \bibinfo {author} {\bibfnamefont {Z.-B.}\ \bibnamefont {Chen}},
  \bibinfo {author} {\bibfnamefont {C.-Y.}\ \bibnamefont {Lu}}, \bibinfo
  {author} {\bibfnamefont {H.}~\bibnamefont {Weinfurter}}, \bibinfo {author}
  {\bibfnamefont {A.}~\bibnamefont {Zeilinger}},\ and\ \bibinfo {author}
  {\bibfnamefont {M.}~\bibnamefont {{\.Z}ukowski}},\ }\bibfield  {title}
  {\bibinfo {title} {Multiphoton entanglement and interferometry},\ }\href
  {https://doi.org/10.1103/RevModPhys.84.777} {\bibfield  {journal} {\bibinfo
  {journal} {Reviews of Modern Physics}\ }\textbf {\bibinfo {volume} {84}},\
  \bibinfo {pages} {777} (\bibinfo {year} {2012})}\BibitemShut {NoStop}%
\bibitem [{\citenamefont {Hillery}\ \emph {et~al.}(1999)\citenamefont
  {Hillery}, \citenamefont {Bu{\v z}ek},\ and\ \citenamefont
  {Berthiaume}}]{hillery_secret_1999}%
  \BibitemOpen
  \bibfield  {author} {\bibinfo {author} {\bibfnamefont {M.}~\bibnamefont
  {Hillery}}, \bibinfo {author} {\bibfnamefont {V.}~\bibnamefont {Bu{\v
  z}ek}},\ and\ \bibinfo {author} {\bibfnamefont {A.}~\bibnamefont
  {Berthiaume}},\ }\bibfield  {title} {\bibinfo {title} {Quantum secret
  sharing},\ }\href {https://doi.org/10.1103/PhysRevA.59.1829} {\bibfield
  {journal} {\bibinfo  {journal} {Physical Review A}\ }\textbf {\bibinfo
  {volume} {59}},\ \bibinfo {pages} {1829} (\bibinfo {year}
  {1999})}\BibitemShut {NoStop}%
\bibitem [{\citenamefont {Christandl}\ and\ \citenamefont
  {Wehner}(2005)}]{christandl_anonymous_2005}%
  \BibitemOpen
  \bibfield  {author} {\bibinfo {author} {\bibfnamefont {M.}~\bibnamefont
  {Christandl}}\ and\ \bibinfo {author} {\bibfnamefont {S.}~\bibnamefont
  {Wehner}},\ }\bibfield  {title} {\bibinfo {title} {Quantum anonymous
  transmissions},\ }in\ \href {https://doi.org/10.1007/11593447_12} {\emph
  {\bibinfo {booktitle} {Lecture Notes in Computer Science}}}\ (\bibinfo
  {publisher} {{Springer Berlin Heidelberg}},\ \bibinfo {year} {2005})\ pp.\
  \bibinfo {pages} {217--235}\BibitemShut {NoStop}%
\bibitem [{\citenamefont {Browne}\ and\ \citenamefont
  {Rudolph}(2005)}]{browne_computation_2005}%
  \BibitemOpen
  \bibfield  {author} {\bibinfo {author} {\bibfnamefont {D.~E.}\ \bibnamefont
  {Browne}}\ and\ \bibinfo {author} {\bibfnamefont {T.}~\bibnamefont
  {Rudolph}},\ }\bibfield  {title} {\bibinfo {title} {Resource-efficient linear
  optical quantum computation},\ }\href
  {https://doi.org/10.1103/PhysRevLett.95.010501} {\bibfield  {journal}
  {\bibinfo  {journal} {Physical Review Letters}\ }\textbf {\bibinfo {volume}
  {95}},\ \bibinfo {pages} {010501} (\bibinfo {year} {2005})}\BibitemShut
  {NoStop}%
\bibitem [{\citenamefont {Bouwmeester}\ \emph {et~al.}(1999)\citenamefont
  {Bouwmeester}, \citenamefont {Pan}, \citenamefont {Daniell}, \citenamefont
  {Weinfurter},\ and\ \citenamefont {Zeilinger}}]{bouwmeester_GHZ_1999}%
  \BibitemOpen
  \bibfield  {author} {\bibinfo {author} {\bibfnamefont {D.}~\bibnamefont
  {Bouwmeester}}, \bibinfo {author} {\bibfnamefont {J.-W.}\ \bibnamefont
  {Pan}}, \bibinfo {author} {\bibfnamefont {M.}~\bibnamefont {Daniell}},
  \bibinfo {author} {\bibfnamefont {H.}~\bibnamefont {Weinfurter}},\ and\
  \bibinfo {author} {\bibfnamefont {A.}~\bibnamefont {Zeilinger}},\ }\bibfield
  {title} {\bibinfo {title} {Observation of three-photon
  {{Greenberger}}-{{Horne}}-{{Zeilinger}} entanglement},\ }\href
  {https://doi.org/10.1103/PhysRevLett.82.1345} {\bibfield  {journal} {\bibinfo
   {journal} {Physical Review Letters}\ }\textbf {\bibinfo {volume} {82}},\
  \bibinfo {pages} {1345} (\bibinfo {year} {1999})}\BibitemShut {NoStop}%
\bibitem [{\citenamefont {Pan}\ \emph {et~al.}(2001)\citenamefont {Pan},
  \citenamefont {Daniell}, \citenamefont {Gasparoni}, \citenamefont {Weihs},\
  and\ \citenamefont {Zeilinger}}]{Pan2001}%
  \BibitemOpen
  \bibfield  {author} {\bibinfo {author} {\bibfnamefont {J.-W.}\ \bibnamefont
  {Pan}}, \bibinfo {author} {\bibfnamefont {M.}~\bibnamefont {Daniell}},
  \bibinfo {author} {\bibfnamefont {S.}~\bibnamefont {Gasparoni}}, \bibinfo
  {author} {\bibfnamefont {G.}~\bibnamefont {Weihs}},\ and\ \bibinfo {author}
  {\bibfnamefont {A.}~\bibnamefont {Zeilinger}},\ }\bibfield  {title} {\bibinfo
  {title} {Experimental demonstration of four-photon entanglement and
  high-fidelity teleportation},\ }\href
  {https://doi.org/10.1103/PhysRevLett.86.4435} {\bibfield  {journal} {\bibinfo
   {journal} {Physical Review Letters}\ }\textbf {\bibinfo {volume} {86}},\
  \bibinfo {pages} {4435} (\bibinfo {year} {2001})}\BibitemShut {NoStop}%
\bibitem [{\citenamefont {Eibl}\ \emph {et~al.}(2003)\citenamefont {Eibl},
  \citenamefont {Gaertner}, \citenamefont {Bourennane}, \citenamefont
  {Kurtsiefer}, \citenamefont {{\.Z}ukowski},\ and\ \citenamefont
  {Weinfurter}}]{Eibl2003}%
  \BibitemOpen
  \bibfield  {author} {\bibinfo {author} {\bibfnamefont {M.}~\bibnamefont
  {Eibl}}, \bibinfo {author} {\bibfnamefont {S.}~\bibnamefont {Gaertner}},
  \bibinfo {author} {\bibfnamefont {M.}~\bibnamefont {Bourennane}}, \bibinfo
  {author} {\bibfnamefont {C.}~\bibnamefont {Kurtsiefer}}, \bibinfo {author}
  {\bibfnamefont {M.}~\bibnamefont {{\.Z}ukowski}},\ and\ \bibinfo
  {author} {\bibfnamefont {H.}~\bibnamefont {Weinfurter}},\ }\bibfield  {title}
  {\bibinfo {title} {Experimental observation of four-photon entanglement from
  parametric down-conversion},\ }\href
  {https://doi.org/10.1103/PhysRevLett.90.200403} {\bibfield  {journal}
  {\bibinfo  {journal} {Physical Review Letters}\ }\textbf {\bibinfo {volume}
  {90}},\ \bibinfo {pages} {200403} (\bibinfo {year} {2003})}\BibitemShut
  {NoStop}%
\bibitem [{\citenamefont {Eibl}\ \emph {et~al.}(2004)\citenamefont {Eibl},
  \citenamefont {Kiesel}, \citenamefont {Bourennane}, \citenamefont
  {Kurtsiefer},\ and\ \citenamefont {Weinfurter}}]{Eibl2004}%
  \BibitemOpen
  \bibfield  {author} {\bibinfo {author} {\bibfnamefont {M.}~\bibnamefont
  {Eibl}}, \bibinfo {author} {\bibfnamefont {N.}~\bibnamefont {Kiesel}},
  \bibinfo {author} {\bibfnamefont {M.}~\bibnamefont {Bourennane}}, \bibinfo
  {author} {\bibfnamefont {C.}~\bibnamefont {Kurtsiefer}},\ and\ \bibinfo
  {author} {\bibfnamefont {H.}~\bibnamefont {Weinfurter}},\ }\bibfield  {title}
  {\bibinfo {title} {Experimental realization of a three-qubit entangled {{W}}
  state},\ }\href {https://doi.org/10.1103/PhysRevLett.92.077901} {\bibfield
  {journal} {\bibinfo  {journal} {Physical Review Letters}\ }\textbf {\bibinfo
  {volume} {92}},\ \bibinfo {pages} {077901} (\bibinfo {year}
  {2004})}\BibitemShut {NoStop}%
\bibitem [{\citenamefont {Zhao}\ \emph {et~al.}(2004)\citenamefont {Zhao},
  \citenamefont {Chen}, \citenamefont {Zhang}, \citenamefont {Yang},
  \citenamefont {Briegel},\ and\ \citenamefont {Pan}}]{Zhao2004}%
  \BibitemOpen
  \bibfield  {author} {\bibinfo {author} {\bibfnamefont {Z.}~\bibnamefont
  {Zhao}}, \bibinfo {author} {\bibfnamefont {Y.}~\bibnamefont {Chen}}, \bibinfo
  {author} {\bibfnamefont {A.}~\bibnamefont {Zhang}}, \bibinfo {author}
  {\bibfnamefont {T.}~\bibnamefont {Yang}}, \bibinfo {author} {\bibfnamefont
  {H.}~\bibnamefont {Briegel}},\ and\ \bibinfo {author} {\bibfnamefont
  {J.}~\bibnamefont {Pan}},\ }\bibfield  {title} {\bibinfo {title}
  {Experimental demonstration of five-photon entanglement and open-destination
  teleportation},\ }\href {https://doi.org/10.1038/nature02643} {\bibfield
  {journal} {\bibinfo  {journal} {Nature}\ }\textbf {\bibinfo {volume} {430}},\
  \bibinfo {pages} {54} (\bibinfo {year} {2004})}\BibitemShut {NoStop}%
\bibitem [{\citenamefont {Walther}\ \emph {et~al.}(2005)\citenamefont
  {Walther}, \citenamefont {Resch}, \citenamefont {Rudolph}, \citenamefont
  {Schenck}, \citenamefont {Weinfurter}, \citenamefont {Vedral}, \citenamefont
  {Aspelmeyer},\ and\ \citenamefont {Zeilinger}}]{Walther2005}%
  \BibitemOpen
  \bibfield  {author} {\bibinfo {author} {\bibfnamefont {P.}~\bibnamefont
  {Walther}}, \bibinfo {author} {\bibfnamefont {K.~J.}\ \bibnamefont {Resch}},
  \bibinfo {author} {\bibfnamefont {T.}~\bibnamefont {Rudolph}}, \bibinfo
  {author} {\bibfnamefont {E.}~\bibnamefont {Schenck}}, \bibinfo {author}
  {\bibfnamefont {H.}~\bibnamefont {Weinfurter}}, \bibinfo {author}
  {\bibfnamefont {V.}~\bibnamefont {Vedral}}, \bibinfo {author} {\bibfnamefont
  {M.}~\bibnamefont {Aspelmeyer}},\ and\ \bibinfo {author} {\bibfnamefont
  {A.}~\bibnamefont {Zeilinger}},\ }\bibfield  {title} {\bibinfo {title}
  {Experimental one-way quantum computing},\ }\href
  {https://doi.org/10.1038/nature03347} {\bibfield  {journal} {\bibinfo
  {journal} {Nature}\ }\textbf {\bibinfo {volume} {434}},\ \bibinfo {pages}
  {169} (\bibinfo {year} {2005})}\BibitemShut {NoStop}%
\bibitem [{\citenamefont {Lu}\ \emph {et~al.}(2007)\citenamefont {Lu},
  \citenamefont {Zhou}, \citenamefont {G{\"u}hne}, \citenamefont {Gao},
  \citenamefont {Zhang}, \citenamefont {Yuan}, \citenamefont {Goebel},
  \citenamefont {Yang},\ and\ \citenamefont {Pan}}]{Lu2007}%
  \BibitemOpen
  \bibfield  {author} {\bibinfo {author} {\bibfnamefont {C.-Y.}\ \bibnamefont
  {Lu}}, \bibinfo {author} {\bibfnamefont {X.-Q.}\ \bibnamefont {Zhou}},
  \bibinfo {author} {\bibfnamefont {O.}~\bibnamefont {G{\"u}hne}}, \bibinfo
  {author} {\bibfnamefont {W.-B.}\ \bibnamefont {Gao}}, \bibinfo {author}
  {\bibfnamefont {J.}~\bibnamefont {Zhang}}, \bibinfo {author} {\bibfnamefont
  {Z.-S.}\ \bibnamefont {Yuan}}, \bibinfo {author} {\bibfnamefont
  {A.}~\bibnamefont {Goebel}}, \bibinfo {author} {\bibfnamefont
  {T.}~\bibnamefont {Yang}},\ and\ \bibinfo {author} {\bibfnamefont {J.-W.}\
  \bibnamefont {Pan}},\ }\bibfield  {title} {\bibinfo {title} {Experimental
  entanglement of six photons in graph states},\ }\href
  {https://doi.org/10.1038/nphys507} {\bibfield  {journal} {\bibinfo  {journal}
  {Nature Physics}\ }\textbf {\bibinfo {volume} {3}},\ \bibinfo {pages} {91}
  (\bibinfo {year} {2007})}\BibitemShut {NoStop}%
\bibitem [{\citenamefont {Yao}\ \emph {et~al.}(2012)\citenamefont {Yao},
  \citenamefont {Wang}, \citenamefont {Xu}, \citenamefont {Lu}, \citenamefont
  {Pan}, \citenamefont {Bao}, \citenamefont {Peng}, \citenamefont {Lu},
  \citenamefont {Chen},\ and\ \citenamefont {Pan}}]{Yao2012}%
  \BibitemOpen
  \bibfield  {author} {\bibinfo {author} {\bibfnamefont {X.-C.}\ \bibnamefont
  {Yao}}, \bibinfo {author} {\bibfnamefont {T.-X.}\ \bibnamefont {Wang}},
  \bibinfo {author} {\bibfnamefont {P.}~\bibnamefont {Xu}}, \bibinfo {author}
  {\bibfnamefont {H.}~\bibnamefont {Lu}}, \bibinfo {author} {\bibfnamefont
  {G.-S.}\ \bibnamefont {Pan}}, \bibinfo {author} {\bibfnamefont {X.-H.}\
  \bibnamefont {Bao}}, \bibinfo {author} {\bibfnamefont {C.-Z.}\ \bibnamefont
  {Peng}}, \bibinfo {author} {\bibfnamefont {C.-Y.}\ \bibnamefont {Lu}},
  \bibinfo {author} {\bibfnamefont {Y.-A.}\ \bibnamefont {Chen}},\ and\
  \bibinfo {author} {\bibfnamefont {J.-W.}\ \bibnamefont {Pan}},\ }\bibfield
  {title} {\bibinfo {title} {Observation of eight-photon entanglement},\
  }\href@noop {} {\bibfield  {journal} {\bibinfo  {journal} {Nature Photonics}\
  }\textbf {\bibinfo {volume} {6}},\ \bibinfo {pages} {225} (\bibinfo {year}
  {2012})}\BibitemShut {NoStop}%
\bibitem [{\citenamefont {Wang}\ \emph {et~al.}(2016)\citenamefont {Wang},
  \citenamefont {Chen}, \citenamefont {Li}, \citenamefont {Huang},
  \citenamefont {Liu}, \citenamefont {Chen}, \citenamefont {Luo}, \citenamefont
  {Su}, \citenamefont {Wu}, \citenamefont {Li}, \citenamefont {Lu},
  \citenamefont {Hu}, \citenamefont {Jiang}, \citenamefont {Peng},
  \citenamefont {Li}, \citenamefont {Liu}, \citenamefont {Chen}, \citenamefont
  {Lu},\ and\ \citenamefont {Pan}}]{Wang2016}%
  \BibitemOpen
  \bibfield  {author} {\bibinfo {author} {\bibfnamefont {X.-L.}\ \bibnamefont
  {Wang}}, \bibinfo {author} {\bibfnamefont {L.-K.}\ \bibnamefont {Chen}},
  \bibinfo {author} {\bibfnamefont {W.}~\bibnamefont {Li}}, \bibinfo {author}
  {\bibfnamefont {H.-L.}\ \bibnamefont {Huang}}, \bibinfo {author}
  {\bibfnamefont {C.}~\bibnamefont {Liu}}, \bibinfo {author} {\bibfnamefont
  {C.}~\bibnamefont {Chen}}, \bibinfo {author} {\bibfnamefont {Y.-H.}\
  \bibnamefont {Luo}}, \bibinfo {author} {\bibfnamefont {Z.-E.}\ \bibnamefont
  {Su}}, \bibinfo {author} {\bibfnamefont {D.}~\bibnamefont {Wu}}, \bibinfo
  {author} {\bibfnamefont {Z.-D.}\ \bibnamefont {Li}}, \bibinfo {author}
  {\bibfnamefont {H.}~\bibnamefont {Lu}}, \bibinfo {author} {\bibfnamefont
  {Y.}~\bibnamefont {Hu}}, \bibinfo {author} {\bibfnamefont {X.}~\bibnamefont
  {Jiang}}, \bibinfo {author} {\bibfnamefont {C.-Z.}\ \bibnamefont {Peng}},
  \bibinfo {author} {\bibfnamefont {L.}~\bibnamefont {Li}}, \bibinfo {author}
  {\bibfnamefont {N.-L.}\ \bibnamefont {Liu}}, \bibinfo {author} {\bibfnamefont
  {Y.-A.}\ \bibnamefont {Chen}}, \bibinfo {author} {\bibfnamefont {C.-Y.}\
  \bibnamefont {Lu}},\ and\ \bibinfo {author} {\bibfnamefont {J.-W.}\
  \bibnamefont {Pan}},\ }\bibfield  {title} {\bibinfo {title} {Experimental
  {{Ten}}-{{Photon Entanglement}}},\ }\href
  {https://doi.org/10.1103/physrevlett.117.210502} {\bibfield  {journal}
  {\bibinfo  {journal} {Physical Review Letters}\ }\textbf {\bibinfo {volume}
  {117}},\ \bibinfo {pages} {210502} (\bibinfo {year} {2016})}\BibitemShut
  {NoStop}%
\bibitem [{\citenamefont {Chen}\ \emph {et~al.}(2017)\citenamefont {Chen},
  \citenamefont {Li}, \citenamefont {Yao}, \citenamefont {Huang}, \citenamefont
  {Li}, \citenamefont {Lu}, \citenamefont {Yuan}, \citenamefont {Zhang},
  \citenamefont {Jiang}, \citenamefont {Peng}, \citenamefont {Li},
  \citenamefont {Liu}, \citenamefont {Ma}, \citenamefont {Lu}, \citenamefont
  {Chen},\ and\ \citenamefont {Pan}}]{Chen2017}%
  \BibitemOpen
  \bibfield  {author} {\bibinfo {author} {\bibfnamefont {L.-K.}\ \bibnamefont
  {Chen}}, \bibinfo {author} {\bibfnamefont {Z.-D.}\ \bibnamefont {Li}},
  \bibinfo {author} {\bibfnamefont {X.-C.}\ \bibnamefont {Yao}}, \bibinfo
  {author} {\bibfnamefont {M.}~\bibnamefont {Huang}}, \bibinfo {author}
  {\bibfnamefont {W.}~\bibnamefont {Li}}, \bibinfo {author} {\bibfnamefont
  {H.}~\bibnamefont {Lu}}, \bibinfo {author} {\bibfnamefont {X.}~\bibnamefont
  {Yuan}}, \bibinfo {author} {\bibfnamefont {Y.-B.}\ \bibnamefont {Zhang}},
  \bibinfo {author} {\bibfnamefont {X.}~\bibnamefont {Jiang}}, \bibinfo
  {author} {\bibfnamefont {C.-Z.}\ \bibnamefont {Peng}}, \bibinfo {author}
  {\bibfnamefont {L.}~\bibnamefont {Li}}, \bibinfo {author} {\bibfnamefont
  {N.-L.}\ \bibnamefont {Liu}}, \bibinfo {author} {\bibfnamefont
  {X.}~\bibnamefont {Ma}}, \bibinfo {author} {\bibfnamefont {C.-Y.}\
  \bibnamefont {Lu}}, \bibinfo {author} {\bibfnamefont {Y.-A.}\ \bibnamefont
  {Chen}},\ and\ \bibinfo {author} {\bibfnamefont {J.-W.}\ \bibnamefont
  {Pan}},\ }\bibfield  {title} {\bibinfo {title} {Observation of ten-photon
  entanglement using thin {{BiB}}\_{{3O}}\_6 crystals},\ }\href
  {https://doi.org/10.1364/optica.4.000077} {\bibfield  {journal} {\bibinfo
  {journal} {Optica}\ }\textbf {\bibinfo {volume} {4}},\ \bibinfo {pages} {77}
  (\bibinfo {year} {2017})}\BibitemShut {NoStop}%
\bibitem [{\citenamefont {Pilnyak}\ \emph {et~al.}(2017)\citenamefont
  {Pilnyak}, \citenamefont {Aharon}, \citenamefont {Istrati}, \citenamefont
  {Megidish}, \citenamefont {Retzker},\ and\ \citenamefont
  {Eisenberg}}]{pilnyak_simple_2017}%
  \BibitemOpen
  \bibfield  {author} {\bibinfo {author} {\bibfnamefont {Y.}~\bibnamefont
  {Pilnyak}}, \bibinfo {author} {\bibfnamefont {N.}~\bibnamefont {Aharon}},
  \bibinfo {author} {\bibfnamefont {D.}~\bibnamefont {Istrati}}, \bibinfo
  {author} {\bibfnamefont {E.}~\bibnamefont {Megidish}}, \bibinfo {author}
  {\bibfnamefont {A.}~\bibnamefont {Retzker}},\ and\ \bibinfo {author}
  {\bibfnamefont {H.~S.}\ \bibnamefont {Eisenberg}},\ }\bibfield  {title}
  {\bibinfo {title} {Simple source for large linear cluster photonic states},\
  }\href {https://doi.org/10.1103/physreva.95.022304} {\bibfield  {journal}
  {\bibinfo  {journal} {Physical Review A}\ }\textbf {\bibinfo {volume} {95}},\
  \bibinfo {pages} {022304} (\bibinfo {year} {2017})}\BibitemShut {NoStop}%
\bibitem [{\citenamefont {Zhong}\ \emph {et~al.}(2018)\citenamefont {Zhong},
  \citenamefont {Li}, \citenamefont {Li}, \citenamefont {Peng}, \citenamefont
  {Su}, \citenamefont {Hu}, \citenamefont {He}, \citenamefont {Ding},
  \citenamefont {Zhang}, \citenamefont {Li}, \citenamefont {Zhang},
  \citenamefont {Wang}, \citenamefont {You}, \citenamefont {Wang},
  \citenamefont {Jiang}, \citenamefont {Li}, \citenamefont {Chen},
  \citenamefont {Liu}, \citenamefont {Lu},\ and\ \citenamefont
  {Pan}}]{zhong_twelve_2018}%
  \BibitemOpen
  \bibfield  {author} {\bibinfo {author} {\bibfnamefont {H.-S.}\ \bibnamefont
  {Zhong}}, \bibinfo {author} {\bibfnamefont {Y.}~\bibnamefont {Li}}, \bibinfo
  {author} {\bibfnamefont {W.}~\bibnamefont {Li}}, \bibinfo {author}
  {\bibfnamefont {L.-C.}\ \bibnamefont {Peng}}, \bibinfo {author}
  {\bibfnamefont {Z.-E.}\ \bibnamefont {Su}}, \bibinfo {author} {\bibfnamefont
  {Y.}~\bibnamefont {Hu}}, \bibinfo {author} {\bibfnamefont {Y.-M.}\
  \bibnamefont {He}}, \bibinfo {author} {\bibfnamefont {X.}~\bibnamefont
  {Ding}}, \bibinfo {author} {\bibfnamefont {W.}~\bibnamefont {Zhang}},
  \bibinfo {author} {\bibfnamefont {H.}~\bibnamefont {Li}}, \bibinfo {author}
  {\bibfnamefont {L.}~\bibnamefont {Zhang}}, \bibinfo {author} {\bibfnamefont
  {Z.}~\bibnamefont {Wang}}, \bibinfo {author} {\bibfnamefont {L.}~\bibnamefont
  {You}}, \bibinfo {author} {\bibfnamefont {X.-L.}\ \bibnamefont {Wang}},
  \bibinfo {author} {\bibfnamefont {X.}~\bibnamefont {Jiang}}, \bibinfo
  {author} {\bibfnamefont {L.}~\bibnamefont {Li}}, \bibinfo {author}
  {\bibfnamefont {Y.-A.}\ \bibnamefont {Chen}}, \bibinfo {author}
  {\bibfnamefont {N.-L.}\ \bibnamefont {Liu}}, \bibinfo {author} {\bibfnamefont
  {C.-Y.}\ \bibnamefont {Lu}},\ and\ \bibinfo {author} {\bibfnamefont {J.-W.}\
  \bibnamefont {Pan}},\ }\bibfield  {title} {\bibinfo {title} {12-{{Photon}}
  entanglement and scalable scattershot boson sampling with optimal
  entangled-photon pairs from parametric down-conversion},\ }\href
  {https://doi.org/10.1103/physrevlett.121.250505} {\bibfield  {journal}
  {\bibinfo  {journal} {Physical Review Letters}\ }\textbf {\bibinfo {volume}
  {121}},\ \bibinfo {pages} {250505} (\bibinfo {year} {2018})}\BibitemShut
  {NoStop}%
\bibitem [{\citenamefont {Greenberger}\ \emph {et~al.}(1990)\citenamefont
  {Greenberger}, \citenamefont {Horne}, \citenamefont {Shimony},\ and\
  \citenamefont {Zeilinger}}]{greenberger_bells_1990}%
  \BibitemOpen
  \bibfield  {author} {\bibinfo {author} {\bibfnamefont {D.~M.}\ \bibnamefont
  {Greenberger}}, \bibinfo {author} {\bibfnamefont {M.~A.}\ \bibnamefont
  {Horne}}, \bibinfo {author} {\bibfnamefont {A.}~\bibnamefont {Shimony}},\
  and\ \bibinfo {author} {\bibfnamefont {A.}~\bibnamefont {Zeilinger}},\
  }\bibfield  {title} {\bibinfo {title} {Bell's theorem without inequalities},\
  }\href {https://doi.org/10.1119/1.16243} {\bibfield  {journal} {\bibinfo
  {journal} {American Journal of Physics}\ }\textbf {\bibinfo {volume} {58}},\
  \bibinfo {pages} {1131} (\bibinfo {year} {1990})}\BibitemShut {NoStop}%
\bibitem [{\citenamefont {H{\"u}bel}\ \emph {et~al.}(2010)\citenamefont
  {H{\"u}bel}, \citenamefont {Hamel}, \citenamefont {Fedrizzi}, \citenamefont
  {Ramelow}, \citenamefont {Resch},\ and\ \citenamefont
  {Jennewein}}]{hubel_direct_2010}%
  \BibitemOpen
  \bibfield  {author} {\bibinfo {author} {\bibfnamefont {H.}~\bibnamefont
  {H{\"u}bel}}, \bibinfo {author} {\bibfnamefont {D.~R.}\ \bibnamefont
  {Hamel}}, \bibinfo {author} {\bibfnamefont {A.}~\bibnamefont {Fedrizzi}},
  \bibinfo {author} {\bibfnamefont {S.}~\bibnamefont {Ramelow}}, \bibinfo
  {author} {\bibfnamefont {K.~J.}\ \bibnamefont {Resch}},\ and\ \bibinfo
  {author} {\bibfnamefont {T.}~\bibnamefont {Jennewein}},\ }\bibfield  {title}
  {\bibinfo {title} {Direct generation of photon triplets using cascaded
  photon-pair sources},\ }\href {https://doi.org/10.1038/nature09175}
  {\bibfield  {journal} {\bibinfo  {journal} {Nature}\ }\textbf {\bibinfo
  {volume} {466}},\ \bibinfo {pages} {601} (\bibinfo {year}
  {2010})}\BibitemShut {NoStop}%
\bibitem [{\citenamefont {Ding}\ \emph {et~al.}(2015)\citenamefont {Ding},
  \citenamefont {Zhang}, \citenamefont {Shi}, \citenamefont {Zhou},
  \citenamefont {Li}, \citenamefont {Shi},\ and\ \citenamefont
  {Guo}}]{ding_hybrid-cascaded_2015}%
  \BibitemOpen
  \bibfield  {author} {\bibinfo {author} {\bibfnamefont {D.-S.}\ \bibnamefont
  {Ding}}, \bibinfo {author} {\bibfnamefont {W.}~\bibnamefont {Zhang}},
  \bibinfo {author} {\bibfnamefont {S.}~\bibnamefont {Shi}}, \bibinfo {author}
  {\bibfnamefont {Z.-Y.}\ \bibnamefont {Zhou}}, \bibinfo {author}
  {\bibfnamefont {Y.}~\bibnamefont {Li}}, \bibinfo {author} {\bibfnamefont
  {B.-S.}\ \bibnamefont {Shi}},\ and\ \bibinfo {author} {\bibfnamefont {G.-C.}\
  \bibnamefont {Guo}},\ }\bibfield  {title} {\bibinfo {title} {Hybrid-cascaded
  generation of tripartite telecom photons using an atomic ensemble and a
  nonlinear waveguide},\ }\href {https://doi.org/10.1364/OPTICA.2.000642}
  {\bibfield  {journal} {\bibinfo  {journal} {Optica}\ }\textbf {\bibinfo
  {volume} {2}},\ \bibinfo {pages} {642} (\bibinfo {year} {2015})}\BibitemShut
  {NoStop}%
\bibitem [{\citenamefont {Krapick}\ \emph {et~al.}(2016)\citenamefont
  {Krapick}, \citenamefont {Brecht}, \citenamefont {Herrmann}, \citenamefont
  {Quiring},\ and\ \citenamefont {Silberhorn}}]{krapick_-chip_2016}%
  \BibitemOpen
  \bibfield  {author} {\bibinfo {author} {\bibfnamefont {S.}~\bibnamefont
  {Krapick}}, \bibinfo {author} {\bibfnamefont {B.}~\bibnamefont {Brecht}},
  \bibinfo {author} {\bibfnamefont {H.}~\bibnamefont {Herrmann}}, \bibinfo
  {author} {\bibfnamefont {V.}~\bibnamefont {Quiring}},\ and\ \bibinfo {author}
  {\bibfnamefont {C.}~\bibnamefont {Silberhorn}},\ }\bibfield  {title}
  {\bibinfo {title} {On-chip generation of photon-triplet states},\ }\href
  {https://doi.org/10.1364/OE.24.002836} {\bibfield  {journal} {\bibinfo
  {journal} {Optics Express}\ }\textbf {\bibinfo {volume} {24}},\ \bibinfo
  {pages} {2836} (\bibinfo {year} {2016})}\BibitemShut {NoStop}%
\bibitem [{\citenamefont {Hamel}\ \emph {et~al.}(2014)\citenamefont {Hamel},
  \citenamefont {Shalm}, \citenamefont {H{\"u}bel}, \citenamefont {Miller},
  \citenamefont {Marsili}, \citenamefont {Verma}, \citenamefont {Mirin},
  \citenamefont {Nam}, \citenamefont {Resch},\ and\ \citenamefont
  {Jennewein}}]{hamel_direct_2014}%
  \BibitemOpen
  \bibfield  {author} {\bibinfo {author} {\bibfnamefont {D.~R.}\ \bibnamefont
  {Hamel}}, \bibinfo {author} {\bibfnamefont {L.~K.}\ \bibnamefont {Shalm}},
  \bibinfo {author} {\bibfnamefont {H.}~\bibnamefont {H{\"u}bel}}, \bibinfo
  {author} {\bibfnamefont {A.~J.}\ \bibnamefont {Miller}}, \bibinfo {author}
  {\bibfnamefont {F.}~\bibnamefont {Marsili}}, \bibinfo {author} {\bibfnamefont
  {V.~B.}\ \bibnamefont {Verma}}, \bibinfo {author} {\bibfnamefont {R.~P.}\
  \bibnamefont {Mirin}}, \bibinfo {author} {\bibfnamefont {S.~W.}\ \bibnamefont
  {Nam}}, \bibinfo {author} {\bibfnamefont {K.~J.}\ \bibnamefont {Resch}},\
  and\ \bibinfo {author} {\bibfnamefont {T.}~\bibnamefont {Jennewein}},\
  }\bibfield  {title} {\bibinfo {title} {Direct generation of three-photon
  polarization entanglement},\ }\href
  {https://doi.org/10.1038/nphoton.2014.218} {\bibfield  {journal} {\bibinfo
  {journal} {Nature Photonics}\ }\textbf {\bibinfo {volume} {8}},\ \bibinfo
  {pages} {801} (\bibinfo {year} {2014})}\BibitemShut {NoStop}%
\bibitem [{\citenamefont {Kim}\ \emph {et~al.}(2006)\citenamefont {Kim},
  \citenamefont {Fiorentino},\ and\ \citenamefont
  {Wong}}]{kim_phase-stable_2006}%
  \BibitemOpen
  \bibfield  {author} {\bibinfo {author} {\bibfnamefont {T.}~\bibnamefont
  {Kim}}, \bibinfo {author} {\bibfnamefont {M.}~\bibnamefont {Fiorentino}},\
  and\ \bibinfo {author} {\bibfnamefont {F.~N.~C.}\ \bibnamefont {Wong}},\
  }\bibfield  {title} {\bibinfo {title} {Phase-stable source of
  polarization-entangled photons using a polarization {{Sagnac}}
  interferometer},\ }\href {https://doi.org/10.1103/PhysRevA.73.012316}
  {\bibfield  {journal} {\bibinfo  {journal} {Physical Review A}\ }\textbf
  {\bibinfo {volume} {73}},\ \bibinfo {pages} {012316} (\bibinfo {year}
  {2006})}\BibitemShut {NoStop}%
\bibitem [{\citenamefont {Fedrizzi}\ \emph {et~al.}(2007)\citenamefont
  {Fedrizzi}, \citenamefont {Herbst}, \citenamefont {Poppe}, \citenamefont
  {Jennewein},\ and\ \citenamefont
  {Zeilinger}}]{fedrizzi_wavelength-tunable_2007}%
  \BibitemOpen
  \bibfield  {author} {\bibinfo {author} {\bibfnamefont {A.}~\bibnamefont
  {Fedrizzi}}, \bibinfo {author} {\bibfnamefont {T.}~\bibnamefont {Herbst}},
  \bibinfo {author} {\bibfnamefont {A.}~\bibnamefont {Poppe}}, \bibinfo
  {author} {\bibfnamefont {T.}~\bibnamefont {Jennewein}},\ and\ \bibinfo
  {author} {\bibfnamefont {A.}~\bibnamefont {Zeilinger}},\ }\bibfield  {title}
  {\bibinfo {title} {A wavelength-tunable fiber-coupled source of narrowband
  entangled photons},\ }\href {https://doi.org/10.1364/OE.15.015377} {\bibfield
   {journal} {\bibinfo  {journal} {Optics Express}\ }\textbf {\bibinfo {volume}
  {15}},\ \bibinfo {pages} {15377} (\bibinfo {year} {2007})}\BibitemShut
  {NoStop}%
\bibitem [{\citenamefont {Lim}\ \emph {et~al.}(2008)\citenamefont {Lim},
  \citenamefont {Yoshizawa}, \citenamefont {Tsuchida},\ and\ \citenamefont
  {Kikuchi}}]{lim_stable_2008}%
  \BibitemOpen
  \bibfield  {author} {\bibinfo {author} {\bibfnamefont {H.~C.}\ \bibnamefont
  {Lim}}, \bibinfo {author} {\bibfnamefont {A.}~\bibnamefont {Yoshizawa}},
  \bibinfo {author} {\bibfnamefont {H.}~\bibnamefont {Tsuchida}},\ and\
  \bibinfo {author} {\bibfnamefont {K.}~\bibnamefont {Kikuchi}},\ }\bibfield
  {title} {\bibinfo {title} {Stable source of high quality telecom-band
  polarization-entangled photon-pairs based on a single, pulse-pumped, short
  {{PPLN}} waveguide},\ }\href {https://doi.org/10.1364/OE.16.012460}
  {\bibfield  {journal} {\bibinfo  {journal} {Optics Express}\ }\textbf
  {\bibinfo {volume} {16}},\ \bibinfo {pages} {12460} (\bibinfo {year}
  {2008})}\BibitemShut {NoStop}%
\bibitem [{\citenamefont {James}\ \emph {et~al.}(2001)\citenamefont {James},
  \citenamefont {Kwiat}, \citenamefont {Munro},\ and\ \citenamefont
  {White}}]{james_measurement_2001}%
  \BibitemOpen
  \bibfield  {author} {\bibinfo {author} {\bibfnamefont {D.~F.~V.}\
  \bibnamefont {James}}, \bibinfo {author} {\bibfnamefont {P.~G.}\ \bibnamefont
  {Kwiat}}, \bibinfo {author} {\bibfnamefont {W.~J.}\ \bibnamefont {Munro}},\
  and\ \bibinfo {author} {\bibfnamefont {A.~G.}\ \bibnamefont {White}},\
  }\bibfield  {title} {\bibinfo {title} {Measurement of qubits},\ }\href
  {https://doi.org/10.1103/PhysRevA.64.052312} {\bibfield  {journal} {\bibinfo
  {journal} {Physical Review A}\ }\textbf {\bibinfo {volume} {64}},\ \bibinfo
  {pages} {052312} (\bibinfo {year} {2001})}\BibitemShut {NoStop}%
\bibitem [{\citenamefont {Coffman}\ \emph {et~al.}(2000)\citenamefont
  {Coffman}, \citenamefont {Kundu},\ and\ \citenamefont
  {Wootters}}]{coffman_distributed_2000}%
  \BibitemOpen
  \bibfield  {author} {\bibinfo {author} {\bibfnamefont {V.}~\bibnamefont
  {Coffman}}, \bibinfo {author} {\bibfnamefont {J.}~\bibnamefont {Kundu}},\
  and\ \bibinfo {author} {\bibfnamefont {W.~K.}\ \bibnamefont {Wootters}},\
  }\bibfield  {title} {\bibinfo {title} {Distributed entanglement},\ }\href
  {https://doi.org/10.1103/PhysRevA.61.052306} {\bibfield  {journal} {\bibinfo
  {journal} {Physical Review A}\ }\textbf {\bibinfo {volume} {61}},\ \bibinfo
  {pages} {052306} (\bibinfo {year} {2000})}\BibitemShut {NoStop}%
\bibitem [{\citenamefont {T{\'o}th}\ and\ \citenamefont
  {G{\"u}hne}(2005)}]{toth_detecting_2005}%
  \BibitemOpen
  \bibfield  {author} {\bibinfo {author} {\bibfnamefont {G.}~\bibnamefont
  {T{\'o}th}}\ and\ \bibinfo {author} {\bibfnamefont {O.}~\bibnamefont
  {G{\"u}hne}},\ }\bibfield  {title} {\bibinfo {title} {Detecting genuine
  multipartite entanglement with two local measurements},\ }\href
  {https://doi.org/10.1103/physrevlett.94.060501} {\bibfield  {journal}
  {\bibinfo  {journal} {Physical Review Letters}\ }\textbf {\bibinfo {volume}
  {94}},\ \bibinfo {pages} {060501} (\bibinfo {year} {2005})}\BibitemShut
  {NoStop}%
\bibitem [{\citenamefont {G{\"u}hne}\ and\ \citenamefont
  {T{\'o}th}(2009)}]{guhne_entanglement_2009}%
  \BibitemOpen
  \bibfield  {author} {\bibinfo {author} {\bibfnamefont {O.}~\bibnamefont
  {G{\"u}hne}}\ and\ \bibinfo {author} {\bibfnamefont {G.}~\bibnamefont
  {T{\'o}th}},\ }\bibfield  {title} {\bibinfo {title} {Entanglement
  detection},\ }\href {https://doi.org/10.1016/j.physrep.2009.02.004}
  {\bibfield  {journal} {\bibinfo  {journal} {Physics Reports}\ }\textbf
  {\bibinfo {volume} {474}},\ \bibinfo {pages} {1} (\bibinfo {year}
  {2009})}\BibitemShut {NoStop}%
\bibitem [{\citenamefont {Birks}\ \emph {et~al.}(1997)\citenamefont {Birks},
  \citenamefont {Knight},\ and\ \citenamefont {Russell}}]{Birks1997}%
  \BibitemOpen
  \bibfield  {author} {\bibinfo {author} {\bibfnamefont {T.~A.}\ \bibnamefont
  {Birks}}, \bibinfo {author} {\bibfnamefont {J.~C.}\ \bibnamefont {Knight}},\
  and\ \bibinfo {author} {\bibfnamefont {P.~S.}\ \bibnamefont {Russell}},\
  }\bibfield  {title} {\bibinfo {title} {Endlessly single-mode photonic crystal
  fiber},\ }\href {https://doi.org/10.1364/OL.22.000961} {\bibfield  {journal}
  {\bibinfo  {journal} {Optics Letters}\ }\textbf {\bibinfo {volume} {22}},\
  \bibinfo {pages} {961} (\bibinfo {year} {1997})}\BibitemShut {NoStop}%
\bibitem [{\citenamefont {Lee}\ \emph {et~al.}(2021)\citenamefont {Lee},
  \citenamefont {Xie}, \citenamefont {Tannous},\ and\ \citenamefont
  {Jennewein}}]{lee_sagnac-type_2021}%
  \BibitemOpen
  \bibfield  {author} {\bibinfo {author} {\bibfnamefont {Y.~S.}\ \bibnamefont
  {Lee}}, \bibinfo {author} {\bibfnamefont {M.}~\bibnamefont {Xie}}, \bibinfo
  {author} {\bibfnamefont {R.}~\bibnamefont {Tannous}},\ and\ \bibinfo {author}
  {\bibfnamefont {T.}~\bibnamefont {Jennewein}},\ }\bibfield  {title} {\bibinfo
  {title} {Sagnac-type entangled photon source using only conventional
  polarization optics},\ }\href {https://doi.org/10.1088/2058-9565/abd151}
  {\bibfield  {journal} {\bibinfo  {journal} {Quantum Science and Technology}\
  }\textbf {\bibinfo {volume} {6}},\ \bibinfo {pages} {025004} (\bibinfo {year}
  {2021})}\BibitemShut {NoStop}%
\bibitem [{\citenamefont {Wang}\ \emph {et~al.}(2018)\citenamefont {Wang},
  \citenamefont {Langrock}, \citenamefont {Marandi}, \citenamefont {Jankowski},
  \citenamefont {Zhang}, \citenamefont {Desiatov}, \citenamefont {Fejer},\ and\
  \citenamefont {Lon{\v c}ar}}]{Wang2018}%
  \BibitemOpen
  \bibfield  {author} {\bibinfo {author} {\bibfnamefont {C.}~\bibnamefont
  {Wang}}, \bibinfo {author} {\bibfnamefont {C.}~\bibnamefont {Langrock}},
  \bibinfo {author} {\bibfnamefont {A.}~\bibnamefont {Marandi}}, \bibinfo
  {author} {\bibfnamefont {M.}~\bibnamefont {Jankowski}}, \bibinfo {author}
  {\bibfnamefont {M.}~\bibnamefont {Zhang}}, \bibinfo {author} {\bibfnamefont
  {B.}~\bibnamefont {Desiatov}}, \bibinfo {author} {\bibfnamefont {M.~M.}\
  \bibnamefont {Fejer}},\ and\ \bibinfo {author} {\bibfnamefont
  {M.}~\bibnamefont {Lon{\v c}ar}},\ }\bibfield  {title} {\bibinfo {title}
  {Ultrahigh-efficiency wavelength conversion in nanophotonic periodically
  poled lithium niobate waveguides},\ }\href
  {https://doi.org/10.1364/optica.5.001438} {\bibfield  {journal} {\bibinfo
  {journal} {Optica}\ }\textbf {\bibinfo {volume} {5}},\ \bibinfo {pages}
  {1438} (\bibinfo {year} {2018})}\BibitemShut {NoStop}%
\end{thebibliography}
\end{document}